\documentclass[twocolumn, tighten, times, twocolappendix]{aastex7}
\usepackage{amsmath}

\begin{document}

\title{BE Lyncis is not a Black Hole Binary:\\Lessons From \textit{Gaia} and \textit{Hipparcos} Astrometry}

\author[orcid=0000-0002-1386-0603,gname=Pranav,sname=Nagarajan]{Pranav Nagarajan}
\affiliation{Department of Astronomy, California Institute of Technology, 1200 E. California Blvd., Pasadena, CA 91125, USA}
\email[show]{pnagaraj@caltech.edu}  

\author[orcid=0000-0002-6871-1752,gname=Kareem,sname=El-Badry]{Kareem El-Badry}
\affiliation{Department of Astronomy, California Institute of Technology, 1200 E. California Blvd., Pasadena, CA 91125, USA}
\email{kelbadry@caltech.edu}  

\author[orcid=0000-0003-0976-4755, gname=Thomas,sname=Maccarone]{Thomas J. Maccarone}
\affiliation{Department of Physics \& Astronomy, Texas Tech University, Box 41051, Lubbock, TX 79409-1051, USA}
\email{thomas.maccarone@ttu.edu}

\author[orcid=0000-0003-0293-503X,gname=Giuliano,sname=Iorio]{Giuliano Iorio}
\affiliation{Institut de Ci\'encies del Cosmos, Universitat de Barcelona, Carrer de Martí i Franquès, 1, 08028 Barcelona, Spain}
\email{giuliano.iorio@icc.ub.edu}  

\author[orcid=0000-0002-5699-5516,gname=Sara,sname=Rastello]{Sara Rastello}
\affiliation{Institut de Ci\'encies del Cosmos, Universitat de Barcelona, Carrer de Martí i Franquès, 1, 08028 Barcelona, Spain}
\email{sara.rastello@icc.ub.edu} 

\author[orcid=0000-0001-9590-3170,gname=Johanna,sname=M\"uller-Horn]{Johanna M\"uller-Horn}
\affiliation{Max Planck Institute for Astronomy, K\"onigstuhl 17, D-69117, Heidelberg, Germany}
\email{mueller-horn@mpia.de}

\begin{abstract}
BE Lyncis (BE Lyn) is a well-studied high-amplitude $\delta$ Scuti variable star (HADS). Recently, \citet{be_lyn_2026} analyzed a 39-year baseline of times of maximum light of BE Lyn, reporting that it is the most eccentric binary known ($e \approx 0.9989$) and hosts the nearest black hole (BH) known to date. We analyze \textit{Hipparcos} and \textit{Gaia} astrometry of BE Lyn, predicting what the observed proper motion anomaly (PMA) over the 25 year baseline between the two missions would be were the companion really a $\gtrsim 17.5\,M_{\odot}$ BH. We find that the predicted PMA is at least an order of magnitude larger than the observed value of $\approx 1.7 \pm 0.8$ mas yr$^{-1}$, regardless of the assumed orientation of the orbit. We  predict the expected \textit{Gaia} DR3 \texttt{RUWE} for different orientations of the putative BH binary, finding that it ranges from $\approx 2.5$--$4.0$, much larger than the reported value of 1.073. The observed value is instead consistent with a low-mass secondary or a single star. We find that BE Lyncis would have received a 7-parameter acceleration solution if it were a BH binary, in contradiction with its absence from the \textit{Gaia} DR3 non-single star catalogs. Finally, we show that the reported orbit is impossible because the luminous star would overflow its Roche lobe at periastron, irrespective of inclination. We recommend caution in interpreting light-travel time effect (LTTE) models that require very high eccentricities, face-on inclinations, or large companion masses. The observed pulsation timing variations are most likely simply a result of red noise or pulsation phase evolution. 
\end{abstract}



\section{Introduction}
\label{sec:intro}

The search for dormant black holes (BH) is in full swing. A handful of these elusive, non-accreting compact objects have been discovered via extensive spectroscopic \citep[e.g.,][]{giesers_detached_2018, giesers_2019, mahy_identifying_2022, shenar_x-ray-quiet_2022},
gravitational microlensing \citep[e.g.,][]{lam_isolated_2022, sahu_isolated_2022}, and astrometric \citep[e.g.,][]{el-badry_sun-like_2023, chakrabarti_noninteracting_2023, tanikawa_2023, el-badry_red_2023, gaia_collaboration_discovery_2024} surveys. There have also been several photometric searches for dormant BHs in close binaries based on the tidal distortion of the luminous companion \citep[ellipsoidal variability; e.g.,][]{rowan_ellipsoidal_2021, green_ellipsoidal_2023, kapusta_mroz_2023, gomel_gaia_2023, nagarajan_spectroscopic_2023}, which have not yet yielded any robust detections.

A complementary search method based on photometry, suitable for dormant BHs with pulsating companions in wide binaries, is measurement of the period modulation of the variable companion due to orbital motion (i.e., the light-travel time effect; \citealt{murphy_ltte_2015, dholakia_tess_2025}). High-amplitude $\delta$ Scuti variable stars (HADS), which display stable periodic signals, photometric variations $\gtrsim 0.3$ mag, and slow rotation, are ideally suited precision ``clocks'' for this task \citep[e.g.,][]{breger_delta_2000}. 

Recently, BE Lyncis (BE Lyn, HIP 45649, \textit{Gaia} DR3 819077377476613888), a well-known HADS, was identified as a dormant BH binary by \citet{be_lyn_2026}. From a $O - C$ analysis of a 39-year baseline of times of maximum light, involving fitting of a model with a intrinsic period variation term and a light-travel time effect (LTTE) term, they derive a best-fit orbital period of $5811$ d and a record-setting best-fit orbital eccentricity of $0.9989$. They infer a best-fit mass of $1.55\,M_{\odot}$ for the luminous star and find an inclination of $i \lesssim 4^{\circ}$ from the requirement that the star should fit inside its Roche lobe at periastron, but as we discuss in Section~\ref{sec:inc}, this constraint is incorrect. 
Together with the orbit inferred from pulsation timing, they infer a minimum companion mass of $M_2 \gtrsim 17.5\,M_{\odot}$. If the model is correct, this would make BE Lyncis ($d = 250$ pc, \citealt{bailer_jones_2021}) the host of the closest known BH to Earth.

Pulsation timing searches are intrinsically biased towards high inclinations, so the low inferred inclination is remarkable. Since astrometric searches have the opposite bias (i.e., towards low inclinations), astrometric measurements from the \textit{Hipparcos} \citep{hipparcos_2007} and \textit{Gaia} \citep{gaia_collaboration_gaia_2016} missions are critically useful in verifying such claims of dormant BH discoveries. A $17.5\,M_{\odot}$ BH would cause the on-sky motion of BE Lyn to deviate significantly from a single-star model, inducing a substantial proper motion anomaly (PMA; e.g., \citealt{kervella_pma_2022}) over the 25-year time baseline between these two missions. Here, we find that the large secondary mass derived by \citet{be_lyn_2026} is in contradiction with both the measured \textit{Hipparcos}--\textit{Gaia} PMA and the excess astrometric noise, or \texttt{RUWE}, value reported in \textit{Gaia}'s third data release (``DR3'', \citealt{gaia_collaboration_gaia_2023}). The data are instead compatible with a higher inclination and a normal stellar companion --- or perhaps even no companion at all --- refuting BE Lyncis as a dormant BH candidate.

The remainder of this paper is organized as follows. In Section~\ref{sec:astrometry}, we analyze the available astrometry of BE Lyn, and show that the DR3 \texttt{RUWE} and \textit{Hipparcos}--\textit{Gaia} proper motion anomaly are incompatible with a dormant BH secondary. In Section~\ref{sec:discussion}, we provide recommendations for vetting future photometric BH candidates. We summarize and conclude in Section~\ref{sec:conclusion}.

\section{Astrometry of BE Lyncis}
\label{sec:astrometry}

Throughout this section, we use the \texttt{gaiamock}\footnote{\texttt{https://github.com/kareemelbadry/gaiamock}} pipeline \citep{el-badry_generative_2024} to forward model \textit{Gaia}'s astrometric observations and model fitting. For a given binary with specified parameters, the pipeline generates mock observations based on the \textit{Gaia} scanning law. The code can also be used to fit the epoch astrometry to determine what kind of solution (i.e., single-star, acceleration, or orbital) the binary will receive. 

We adopt the best-fit orbital parameters reported in \citet{be_lyn_2026}. We retrieve the right ascension (RA or $\alpha$), declination (Dec or $\delta$), parallax, and mean $G$-band apparent magnitude of BE Lyn from the DR3 catalog. We adopt center-of-mass proper motions such that the predicted proper motions in DR3 match the observed values to within the reported uncertainties. Since the longitude of the ascending node of BE Lyn is unknown, we assume $\Omega = 60^{\circ}$ for illustrative purposes; we consider the effects of varying $\Omega$ in Section~\ref{sec:Omega}.

\subsection{\textit{Gaia} DR3 \texttt{RUWE}}
\label{sec:ruwe}

The observed DR3 \texttt{RUWE} for BE Lyn is 1.073, well under the nominal binarity threshold of $\texttt{RUWE} = 1.4$. Using \texttt{gaiamock}, we find that the DR3 \texttt{RUWE} would be closer to $\approx 3.9$ for a dark, $\gtrsim 17.5\,M_{\odot}$ companion. We have shown previously that \texttt{gaiamock} reliably predicts \texttt{RUWE} values for binaries with detectable astrometric orbits \citep{el-badry_inflation_2025, Muller-Horn2025}.  A \texttt{RUWE} value of $\approx 3.9$ implies that the astrometric residuals are on average $\approx 3.9$ times larger than expected, implying a very poor fit. In contrast, a \texttt{RUWE} of 1.07 implies a good fit with a single-star model \citep[e.g.,][]{belokurov_ruwe_2020}. Hence, the predicted DR3 \texttt{RUWE} is inconsistent with the observed DR3 \texttt{RUWE}. On the other hand, a larger inclination and a much lower companion mass are compatible with the observed DR3 \texttt{RUWE}, implying that the secondary could be a low-mass dwarf. 

\subsection{Proper Motion Anomaly}
\label{sec:pma}

For BE Lyn, the proper motion in RA is reported to be $9.29 \pm 0.88$ mas yr$^{-1}$, $7.878 \pm 0.071$ mas yr$^{-1}$, and $7.950 \pm 0.020$ mas yr$^{-1}$ in the new \textit{Hipparcos} reduction, \textit{Gaia} DR2, and \textit{Gaia} DR3, respectively \citep{hipparcos_2007, brown_dr2_2018, gaia_collaboration_gaia_2023}. In addition, the proper motion in Dec is reported to be  $1.87 \pm 0.61$ mas yr$^{-1}$, $0.521 \pm 0.075$ mas yr$^{-1}$, and $0.753 \pm 0.018$ mas yr$^{-1}$ in \textit{Hipparcos}, DR2, and DR3, respectively \citep{hipparcos_2007, brown_dr2_2018, gaia_collaboration_gaia_2023}. The small difference between these values represents at best marginal evidence for binary motion \citep[e.g.,][]{brandt_accelerations_2021}.

\begin{figure*}[t!]
    \centering
    \includegraphics[width=\textwidth]{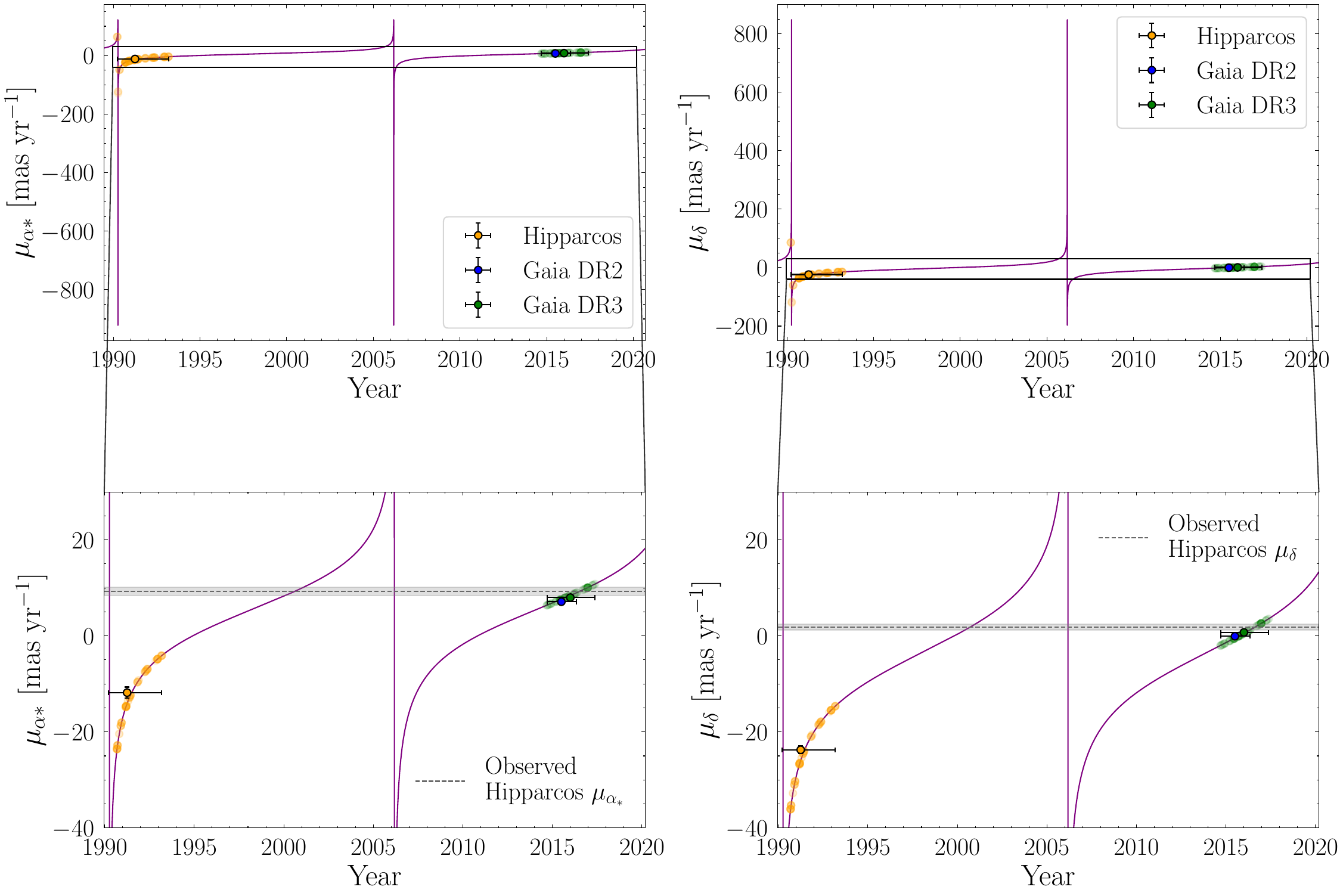}
    \caption{Predicted instantaneous proper motions for BE Lyn, assuming the best-fit orbital parameters reported by \citet{be_lyn_2026}. We adopt $\Omega = 60^{\circ}$. We show the predicted \textit{Hipparcos}, \textit{Gaia} DR2, and \textit{Gaia} DR3 proper motions as well, with the predicted individual epoch measurements plotted with lower opacity. We zoom in on these astrometric measurements in the lower panels. In these panels, we denote the measured \textit{Hipparcos} proper motions with dashed lines; the shaded regions represent the corresponding uncertainties. Clearly, the observed and predicted \textit{Hipparcos} proper motions are quite different. We predict that we would observe a proper motion anomaly of $\Delta \mu_{\alpha*} \approx 20$ mas yr$^{-1}$ and  $\Delta \mu_{\delta} \approx 24$ mas yr$^{-1}$ over a 25-year baseline, far greater than the $1$--$2$ mas yr$^{-1}$ differences that are actually observed.}
    \label{fig:proper_motions}
\end{figure*}

We use the best-fit orbital parameters to analytically compute the instantaneous proper motions of BE Lyn, were it a dormant BH binary as claimed. In detail, in the center-of-mass frame:

\begin{equation}
    \mu_{\alpha *} = A \dot{X} + F\dot{Y},\,\,
    \mu_{\delta} = B \dot{X} + G\dot{Y}
\end{equation}

\noindent where $A, B, F,$ and $G$ are the Thiele-Innes coefficients, 

\begin{equation}
    X = \cos{E} - e,\,\, Y = \sqrt{1 - e^2} \sin E
\end{equation}

\noindent are the coordinates of the primary in the orbital plane, and $e$ and $E$ are the eccentricity and eccentric anomaly, respectively \citep[e.g.,][]{bennendijk_1960}. We plot these instantaneous proper motions across the 25-year baseline between \textit{Hipparcos} and \textit{Gaia} in Figure~\ref{fig:proper_motions}. Due to the binary's extreme eccentricity, the variability of the proper motions is sharply peaked about periastron, with an amplitude between the minimum and maximum values of $\sim 1000$ mas yr$^{-1}$.

We retrieve the individual \textit{Hipparcos} epochs, scan angles, and parallax factors for BE Lyn using the Intermediate Astrometric Data tool \citep{brandt_htof_2021}. Using that data, we predict the epoch astrometry that \textit{Hipparcos} would measure for the putative BH binary at those times, and fit those measurements with a single-star astrometric model to derive the expected \textit{Hipparcos} proper motions.\footnote{To do this, we follow the approach detailed in Sections 3.4 and 4.1 of \citet{el-badry_generative_2024}.} In addition, we use \texttt{gaiamock} to predict epoch astrometry for the putative BH binary in \textit{Gaia} DR2\footnote{The reference epoch for \textit{Gaia}'s second data release, or DR2, is J2015.5 = JD 2457206.375. The epoch astrometry for DR2 only extends until JD = 2457532.} and DR3, and fit those measurements with a single-star astrometric model to predict the reported proper motions in those data releases. We show these predicted proper motions in Figure~\ref{fig:proper_motions}, with the individual epoch measurements  plotted with lower opacity. We denote the observed \textit{Hipparcos} proper motions with dashed lines in the lower panels of Figure~\ref{fig:proper_motions}, with the shaded regions representing the corresponding uncertainties. Despite the fact that we adopted center-of-mass proper motions that make the predicted and observed DR3 proper motions agree, the predicted and observed \textit{Hipparcos} proper motions in RA and Dec differ by $\approx 21$ mas yr$^{-1}$ and $\approx 25$ mas yr$^{-1}$, respectively.

We find that, were BE Lyn a BH binary, we would observe a \textit{Hipparcos}--\textit{Gaia} PMA of $\lvert \Delta \mu_{\alpha*} \rvert \approx 20 \pm 1$ mas yr$^{-1}$ and  $\lvert \Delta \mu_{\delta} \rvert \approx 24 \pm 1$ mas yr$^{-1}$ over the 25-year baseline between the two missions. This is an order of magnitude greater than the $\approx 1$--$2$ mas yr$^{-1}$ proper motion differences that are actually observed. In addition, we predict the total PMA between DR2 and DR3 to be $\lvert \Delta \mu \rvert = \left\lvert \sqrt{\Delta \mu_{\alpha*}^2 + \Delta \mu_{\delta}^2}\right\rvert = 1.20 \pm 0.07$ mas yr$^{-1}$. However, the observed PMA between DR2 and DR3 is only $0.24 \pm 0.08$ mas yr$^{-1}$, inconsistent with the expected value by $\approx9\sigma$.

\subsection{\textit{Gaia} DR3 Non-Single Star Catalog}

BE Lyn is not in the \textit{Gaia} DR3 non-single star catalogs; that is, it did not receive an astrometric binary solution. Using \texttt{gaiamock}, we find that BE Lyn would be expected to receive a 7-parameter acceleration solution in DR3 if it harbored a $\gtrsim 17.5\,M_{\odot}$ dormant BH. This conclusion remains unchanged even if additional astrometric scatter due to stellar variability is taken into account \citep{Giuliano26}.

\subsection{Longitude of the Ascending Node}
\label{sec:Omega}

While we adopt a value for the longitude of the ascending node $\Omega$ of BE Lyn of $60^{\circ}$, our broad conclusions are not sensitive to the assumed value of $\Omega$. To test this, we simulate 100 random values of $\Omega$, and repeat our calculation of the \textit{Gaia} DR3 \texttt{RUWE} and the \textit{Hipparcos}-\textit{Gaia} DR3 PMA. We find that the predicted DR3 \texttt{RUWE} varies between $\approx 2.5$ and $\approx 4.0$, consistently larger than the observed value. Furthermore, we find that the total proper motion difference varies between $22$ mas yr$^{-1}$ and $33$ mas yr$^{-1}$ depending on the value of $\Omega$. Clearly, even the minimum value is significantly larger than the observed proper motion anomaly.

\section{Discussion} 
\label{sec:discussion}

\subsection{Reported inclination constraint}
\label{sec:inc}
\citet{be_lyn_2026} report (their Figure 3) that the orbital inclination must be $i \lesssim 4^{\circ}$, because otherwise the luminous star would overflow its Roche lobe at periastron. We are unable to reproduce their calculation and find that the star {\it does} significantly overflow its Roche lobe at periastron for their best-fit orbit when the Roche lobe effective radius is calculated using the fitting function from \citet{eggleton_roche_1983}. In fact, the requirement that the star should not overflow its Roche lobe at periastron does not lead to any significant constraint on the orbital inclination, because the inclination-dependence in $a$ and in the Eggleton factor $f_q$ almost exactly cancel. The effective Roche lobe radius of the primary ranges between $0.79\,R_{\odot}$ and $0.98\,R_{\odot}$ at periastron for the provided best-fit orbital parameters and inclinations ranging from 0$^{\circ}$ to 90$^{\circ}$. While the Eggleton formula and Roche model may not be suitable for such a highly eccentric binary, the fitting formulae for non-synchronous, eccentric binaries from \citet{sepinsky_2007} result in volume-equivalent Roche lobe radii that differ by only $\lesssim 10\%$. 

We conjecture that the contradictory result found by \citet{be_lyn_2026} is a result of using $q=M_2/M_1$ instead of $q=M_1/M_2$ in the Eggleton formula. Indeed, while this paper was undergoing peer review, \citet{be_lyn_2026} corrected this error, relaxing their inclination constraint to $i \leq 10.1^{\circ}$ and their companion mass constraint to $M_2 \geq 2.5\,M_{\odot}$ based on the fact that the periastron separation must be greater than the best-fit radius of the luminous star. They acknowledge that the luminous star dramatically overflows its Roche lobe at periastron, but speculate that this periastron mass transfer led to the wide, extremely eccentric orbit that we observe today.

We consider this revised scenario to be problematic for the following reasons. First, using \texttt{gaiamock}, we find the predicted \textit{Hipparcos}--\textit{Gaia} PMA for $i = 10.1^{\circ}$ and $M_2 = 2.5\,M_{\odot}$ to be $\lvert \Delta \mu_{\alpha*} \rvert \approx 8 \pm 1$ mas yr$^{-1}$ and $\lvert \Delta \mu_{\delta} \rvert \approx 10 \pm 1$ mas yr$^{-1}$. These values are still significantly larger than the $\approx 1$--$2$ mas yr$^{-1}$ proper motion differences that are actually observed. Furthermore, if the binary hosted a $\gtrsim 2.5\,M_{\odot}$ dark companion, it would still have been expected to receive a 7-parameter acceleration solution in DR3. Finally, even a $2.5\,M_{\odot}$ BH companion would tidally disrupt the luminous star at periastron, since the periastron separation is predicted to be smaller than the tidal disruption radius of $R_1 \left(M_2 /M_1\right)^{1/3}$ at all inclinations $i < 10.1^{\circ}$. Such a close encounter could have led to a micro-tidal disruption event \citep[e.g.,][]{perets_mtde_2016}.

\subsection{Recommendations for vetting dormant BH candidates} 

BE Lyncis joins a growing list of BH impostors with inferred inclinations that are close to face-on. As belabored by \citet{el-badry_ngc_2022-1}, when considering such candidates, two caveats should be emphasized: (a) dormant BH binaries are an intrinsically rare population, and (b) low inclinations are intrinsically rare, with sufficiently low inclinations enough to turn any binary into a BH candidate. Indeed, low inclinations are geometrically disfavored, and photometric surveys based on either ellipsoidal variability or the light-travel time effect are biased toward high inclinations. 

We advocate approaching such BH candidates with a healthy dose of skepticism, especially if a higher inclination would make the data consistent with a less remarkable secondary. Since astrometric searches have an inverse bias relative to photometric or spectroscopic searches, it is useful to make use of available \textit{Hipparcos} and \textit{Gaia} measurements before making extraordinary claims. The development of tools such as \texttt{gaiamock}, which forward-models the DR3 astrometric pipeline, makes it straightforward to use metrics such as \texttt{RUWE} to verify whether these measurements are consistent with the presence of a dormant BH. 

Moreover, finding an extremely high eccentricity when fitting a Keplerian orbit is a rather common sign that something has gone wrong. While we are not in a position to say definitively what caused the spurious detection of an LTTE orbit reported by \citet{be_lyn_2026}, we do not find the arrival time periodicity found in their Figure 1 very convincing. We conjecture that it is simply the result of red noise or changes in light curve shape due to, for instance, nonlinear mode interactions \citep[e.g.,][]{breger_delta_2000}.

\section{Conclusion}
\label{sec:conclusion}

From an $O - C$ analysis of times of maximum light over a 39-year baseline, \citet{be_lyn_2026} find that the high-amplitude $\delta$ Scuti variable star BE Lyncis is an extremely eccentric binary that hosts the nearest known dormant black hole (BH) to date. By analyzing available \textit{Gaia} and \textit{Hipparcos} astrometry, we refute this claim. We summarize our conclusions below.

\begin{itemize}
    \item We predict what the observed proper motion anomaly (PMA) over the 25 year baseline between \textit{Gaia} and \textit{Hipparcos} would be were the companion really a $\gtrsim 17.5\,M_{\odot}$ BH. We find that the total predicted PMA ranges between $22$ mas yr$^{-1}$ and $33$ mas yr$^{-1}$ depending on the adopted value of the the longitude of the ascending node $\Omega$. This is an order of magnitude larger than the observed PMA of $1.7 \pm 0.8$ mas yr$^{-1}$, ruling out the proposed model.
    \item We also predict the expected DR3 \texttt{RUWE} for the putative BH binary for different values of $\Omega$, finding that it ranges from $\approx 2.5$--$4.0$, much larger than the reported value of 1.073. This measurement is instead consistent with larger inclinations and a low-mass secondary, or a single star. 
    \item We are unable to reproduce the dynamical constraint on the orbital inclination and companion mass reported by \citet{be_lyn_2026}, even in the case where the LTTE orbit is correct (Section~\ref{sec:inc}). We find that the reported orbit would lead to tidal disruption of the star for any inclination. 
    
    \item Since photometric searches for dormant BHs in binaries are biased toward high inclinations, we remind readers of the importance of caution in interpreting photometric models that require nearly face-on inclinations and large companion masses. This is especially true when these results are in tension with existing astrometric measurements, since astrometric searches have the opposite inclination bias.
\end{itemize}

Our results show that the light-travel time effect (LTTE) model proposed by \citet{be_lyn_2026} for BE Lyn is probably incorrect. Long-term spectroscopic monitoring of BE Lyn, interferometric measurements, or epoch astrometry from \textit{Gaia} DR4 is likely required to reveal the true nature of any binary companion.

\begin{acknowledgments}
We benefited from discussions at the ``Stellar-Mass Black Holes at the Nexus of Optical, X-ray, and Gravitational Wave Surveys'' program at the Kavli Institute of Theoretical Physics. This research was supported by NSF grant AST-2307232. This research was supported in part by grant NSF PHY-2309135 to the Kavli Institute for Theoretical Physics (KITP). This work has made use of data from the European Space Agency (ESA) mission {\it Gaia} (\url{https://www.cosmos.esa.int/gaia}), processed by the {\it Gaia}
Data Processing and Analysis Consortium (DPAC,
\url{https://www.cosmos.esa.int/web/gaia/dpac/consortium}). Funding for the DPAC has been provided by national institutions, in particular the institutions
participating in the {\it Gaia} Multilateral Agreement. 
\end{acknowledgments}

\begin{contribution}

PN was responsible for performing the data analysis and writing the paper. KE came up with the initial research concept and obtained the funding. TM contributed valuable discussion on the \textit{Hipparcos}-\textit{Gaia} proper motion anomaly. GI and SR provided insight on the effect of variability on \textit{Gaia}'s astrometric model fitting. JMH investigated the dynamical stability of the system at periastron.


\end{contribution}

%

\software{astropy \citep{2013A&A...558A..33A, 2018AJ....156..123A, 2022ApJ...935..167A}}





\bibliography{bibliography}{}
\bibliographystyle{aasjournalv7_modified}



\end{document}